\documentclass [12pt,preprint]{aastex}

\begin{document}

\title{A Model for Dark Matter Halos}

\author{F.D.A. Hartwick}

\affil{Department of Physics and Astronomy, \linebreak University of 
Victoria,
Victoria, BC, Canada, V8W 3P6}
\begin {abstract}

A halo model is presented which possesses a constant phase space 
density (Q) core followed by a radial CDM-like power law decrease in Q. 
The motivation for the core is the allowance for a possible primordial phase 
space density limit such as the Tremaine-Gunn upper bound. 
The space density profile derived from this 
model has a constant density core and falls off rapidly beyond.
The new model is shown to improve the fits to the observations of LSB galaxy 
rotation curves, naturally provides a model which has been shown to result 
in a lengthened dynamical friction time scale for the Fornax dwarf
spheroidal galaxy and predicts a flattening of the density profile within 
the Einstein radius of galaxy clusters. A constant gas entropy floor is 
predicted whose adiabatic constant provides a lower limit in accord with 
observed galaxy cluster values. While `observable-sized' cores are not seen 
in standard cold dark matter (CDM) simulations, phase space considerations 
suggest that they could appear in warm dark matter (WDM) cosmological 
simulations and in certain hierarchically consistent SuperWIMP scenarios. 

\end{abstract}


\section {Introduction}

Simulations of cold dark matter (CDM) cosmology predict halos whose 
density profiles are generally well described by what has become known 
as an NFW 
profile (Navarro et al. 1997). A defining characteristic is the presence
of a density cusp at the center. Another important property of CDM halo
profiles is that they exhibit a power law in the parameter $\rho/\sigma^{3}$
which extends
over two orders of magnitude in radius beyond the resolution limit of
the simulations. This was first shown by Taylor \& Navarro (2001) from  
modelling based on the NFW density profile and later directly from CDM  
simulations by Dehnen \& McLaughlin (2005). Nonparametric models of the 
density profiles of CDM halos as well as alternate parameterizations are given 
by Merritt et al. (2006). In a companion paper Graham et al. (2006) discuss 
the power law nature of the $\rho/\sigma^{3}$ profile.

The NFW profile has been successfully fit to many observations including 
those of dwarf spheroidal and low
surface brightness (LSB) galaxies as well as galaxy clusters. However,
not all LSB rotation curves can be fit (Hayashi et al. 2004), and some
authors have reported density profiles near the centers of galaxy clusters
which are shallower than predicted by NFW (e.g. Sand et al. 2002, Broadhurst 
et al. 2005). Recently, Goerdt et al. (2006) have argued that a constant
density core is required in the Fornax dwarf in order to understand why
its resident globular clusters have not disappeared due dynamical friction,
as might be expected if its dark matter halo was of NFW form. Modelling 
dark matter halos with constant density cores is not new but it is usually 
done by parameterizing the density profile (e.g. see Burkert, 1995 and 
references above). Here the goal is to follow the effects of
a finite primordial phase space density upper limit. Thus the constant density 
core results from a solution of the Jeans equation with a parameterized phase 
space density profile.

Simple analytical arguments suggest that the effects of a primordial 
phase space density bound should be seen in present 
structures even after many mergings (e.g. Dalcanton \& Hogan, 2001). In the 
absence of cosmological simulations which include such a primordial bound, 
we rely here on the good agreement of predictions from a simple model 
with observations to argue that standard CDM simulations and hence the NFW 
profile may not be giving a complete picture.

\section{The Model}

We start with a CDM like power law in the quantity $\rho/\sigma^{3}$ where 
$\rho$ is the local space density and $\sigma$ is the local radial velocity 
dispersion. The above quantity is often loosely referred to as
the phase space density (as it is here) but it is actually a `pseudo' phase 
space density (e.g. see Dekel $\&$ Arad (2004) for a discussion of the true 
6-D phase space density). This power law is maintained in the outer regions 
of the model but with a continued rise at  
sufficiently small radius, the phase space density is assumed to reach the 
Tremaine \& Gunn (1979) limit in the absence of other lower and less 
fundamental limiting effects. This quantum statistical upper limit on the 
phase space density applies to thermal particles as well as fermions. Hence, 
as long as the particles are not bosons we expect an eventual
cap/core in the central phase space density. Here a simple model is proposed 
in order to mimic
the lingering effects of a putative but as yet unknown primordial phase space
density 
bound (Q$_{p}\equiv\bar{\rho}/{\bar{\sigma}^{3}}$ where $\bar{\rho}$ is the 
mean density and $\bar{\sigma}$ is the one dimensional velocity dispersion). 
With a constant 
central phase space density ($Q_{o}$) core of size r$_{c}$, the following 
profile is defined:

\begin{equation}
\rho/\sigma^{3}\equiv  Q(r) = \frac{Q_{o}}{(1+(r/r_{c})^{\alpha})^{1.92/\alpha
}}
\end{equation}

The choice of a model independent power law index of 1.92 (close to that 
found by Taylor \& Navarro) comes 
from the work of Dehnen \& McLaughlin (2005). All but one of the models 
presented here were computed with the shape parameter $\alpha=1.92$. Lower 
values of $\alpha$ 
result in a more gradual transition to the outer power law and as will be 
shown produce very similar results.

The above expression for Q(r) is substituted into the Jeans equation  
after eliminating $\sigma^{2}$, thereby allowing the determination of the 
density profile $\rho$. (Note the assumption of spherical symmetry).

\begin{equation}
\frac{d\log \ \rho}{d\log \ r} = -0.6\frac{GM_{r}}{r}\left(\frac{Q}{\rho}
\right)^{2/3}-1.2\beta +0.4\frac{d\log \ Q}{d\log \ r}
\end{equation}

 with 

\begin{equation}
\frac{d M_{r}}{d\log \ r} = \ln(10)  4  \pi  r^{3}  \rho
\end{equation}
where $\beta$ is the anisotropy parameter (Binney \& Tremaine, 1987) and is 
positive for a predominantly radial anisotropy. The following dimensionless 
number involving the initial conditions was found to lie between 2 and 3 for 
all models presented here. 

\begin{equation}
\gamma = 4\pi G r^{2}_{c} Q_{o}(\sigma^{2}_{o})^{1/2}=4\pi G r^{2}_{c} Q^{2/3}
_{o}\rho^{1/3}_{o}
\end{equation}

Integrations can be carried out with $\beta = 0$, but for the above range of 
$\gamma$, the dispersion is found to 
increase outwards around the point of gradient change in Q(r). As CDM 
simulations indicate a small radial anisotropy in the $outer$ parts of a  
halo, at each step in the integration,  
trial values of $\beta$ are stepped through (in units of 0.001) in order 
to determine that value which makes the logarithmic gradient in 
dispersion have the shallowest $negative$ value. In this way a $\beta$ 
profile is obtained starting at zero in the center (actually 0.001 for 
computational reasons) remaining small throughout the core
and usually ending up $\sim0.2-0.3$. At some point 
further out the scheme 
tries to make $\beta$ decrease but it is constrained to remain at its 
maximum value. The above value of $\beta_{max}$ is a rough average of the 
outermost 
values determined for CDM halos (Fig. 3 of Dehnen \& McLaughlin (2005)). 
Unlike models computed without the above simple, well defined 
prescription, those here exhibit consistent scaling relations (see discussion 
below). Furthermore, the initial rise 
of $\beta$ is very nearly a linear function of the logarithmic density 
gradient as advocated
by Hansen \& Moore (2006). Beyond $\beta\sim0.3$ however, both the Hansen \& 
Moore and the Dehnen \& McLaughlin results show a large scatter in $\beta$ and
we have elected to keep it constant in this region. 

The core can be 
considered isothermal (i.e. constant velocity dispersion and negligible 
velocity anisotropy) within the region where the logarithmic density 
gradient is greater than -0.1. This radius is $\sim0.2$~r$_{c}$. 

Depending on what observations are given (i.e. initial rotation curve slope, 
the location of the bend in the rotation curve or its amplitude) determines 
which of the parameters $\rho_{o}$, r$_{c}$ or Q$_{o}$ one chooses to fix  
initially. A model is constructed by integrating equations 
(2) and (3) while systematically varying the other two parameters until 
the logarithmic density gradient becomes -4.000 at M$_{vir}$\footnote{
Approximate solutions which obviate the need for the trial and error procedure
are given in the appendix.}.
M$_{vir}$ and R$_{vir}$ are 
defined as the values of mass and radius where the mean density  becomes  
$100\times\rho_{c}$ (h$_{100}=0.7$ assumed throughout).  This outer boundary 
condition was determined empirically (e.g. by fitting the outer profile to the
observations of the cluster A1689 as described below). The density profile 
derived from equation (2) is quite different from an NFW profile. It possesses
a constant density core followed by a relatively steep radial fall-off. A 
steep fall-off in density appears to be demanded by the observations of galaxy
cluster A1689 (Broadhurst et al. 2005) and is shown in Fig. 1. The value of 
$\rho_{o}$ 
determined for a converged model corresponds 
to a particular value of $\beta_{max}$. Remarkably, any other model with 
{\it{the same}} $\rho_{o}$ can then  
be obtained from the following scaling relations (i.e. Q$_{o}\propto M^
{-1}$, Q$_{o}\propto \sigma_{o}^{-3}$ and Q$_{o}\propto r^{-3}_{c}$) and hence 
is a member of a {\it{one parameter}} family of models. Interestingly,
these scaling relations are identical to those discussed by Dalcanton \& Hogan 
(2001) to describe the results of `gentle' merging given that during a merger 
Q cannot increase. Based on this discussion,  
the inverse relation between Q$_{o}$ and M found here suggests that equation 
(1), for all of its simplicity, is consistent with {\it{a form of}} 
hierarchical structure 
formation. As discussed by the above authors, it is decidedly not compatible 
with `phase packing' where one expects more massive objects (with higher 
central velocity dispersions) to have smaller core radii.  

In what follows, models are specified by the four parameters Q$_{o}$ 
(M$_{\odot}pc^{-3}(km \ sec^{-1})^{-3}$), r$_{c}$ (kpc), $\rho_{o}$ 
(M$_{\odot}pc^{-3}$) and $\alpha$ and are enclosed in brackets.

It is useful to express the equivalent {\it{gas}} entropy 
in terms of Q. We do this by evaluating the adiabatic constant $K=kTn_{e}^
{-2/3}$. With Q in the same units as above 
we obtain 

\begin{equation}
K=\frac{8.84\times10^{-7}\mu\mu_{e}^{2/3}((3-2\beta(r))/3)}{Q^{2/3}}=K_{o}
(3-2\beta(r))/3)(1+(r/r_{c})^{\alpha})^{1.28/\alpha}
\end{equation}
and $K_{o}=8.84\times10^{-7}\mu\mu_{e}^{2/3}Q_{o}^{-2/3}$ kev cm$^{-2}$

Note that the entropy of the gas is initially constant at $K_{o}$ and then 
increases as a power law with index 1.28 as long as $\beta$ remains constant 
in the outer region (see Fig. 1). We emphasize that  
inherent in equation (5) is the assumption that the entropy of the gas is the 
same as the entropy of the dark matter and that as merging continues the 
increase in entropy is the same for both components. Thus this value of K 
must be a lower limit to the actual gas entropy and as such it provides a 
floor on which gas physics processes (i.e. cooling, heating, astration 
etc.) can be played out.

The characteristics of a representative model (in this case for the galaxy 
cluster A1689) are shown in Fig. 1. 

\section{Confronting the Model with Observations}

As a test of the model we apply it to three regimes of total mass: LSB
galaxies, clusters of galaxies and dwarf spheroidal galaxies.

\subsection{LSB Galaxies}

Hayashi et al. (2004) have derived best fit rotation curves for a sample 
of LSB galaxies using the NFW density profile. They divided their fits into
three categories. The first (A class) provided good fits to the observations. 
The second (B class) included galaxies which could not be satisfactorily fit 
with $\Lambda$CDM-compatible parameters. Galaxies in the third group (C class) 
have irregular rotation curves. Fig. 2 shows the fits of the circular velocity
(GM$_{r}/r)^{1/2}$ of our model to four of
the galaxies investigated by Hayashi et al. The observational data is from 
McGaugh et al. 2001 and is available at http://www.astro.umd.edu/~ssm/data. 
Following Hayashi et al. we let the smallest uncertainty in velocity be 
$\pm 5~ kmsec^{-1}$. The above model provides 
adequate fits to both class A $and$ class B samples. Parameters for  the class
A galaxies ($3.41\times10^{-7},1.1,0.254,1.92$) for ESO2060140 and ($1.2\times
10^{-7}
,1.83,0.110,1.92$) for F563-1 show significantly higher central densities 
(and $\beta_{max}$'s$\sim0.3$) 
than the B group galaxies ($3.5\times10^{-8},5.1,5.71\times10^{-3},1.92$) for
ESO0840411 and ($1.68\times10^{-8},6.0,8.28\times10^{-3},1.92$) for UGC5750 
with $\beta_{max}$'s 0.205 and 0.220 
respectively. Generally, models with higher values of $\rho_{o}$ have cores 
with relatively larger values of Q$_{o}$. 

\subsection{Clusters of Galaxies}

The derived behavior of the dark matter density profile in the inner parts of 
galaxy 
clusters is controversial in part because of `contamination' by the baryonic
component in addition to the observational resolution difficulties. Here we
fit our model to a recent gravitational lensing study of the massive cluster 
A1689 by Broadhurst et al. (2005). Standard integration of the density 
profile provides the run of projected mass with radius. Figure 1 shows a model
with parameters ($2.08\times10^{-11},33.4,0.10,1.92$) fit to the
Broadhurst et al. data. The model has a virial
mass 
of $1.3\times10^{15} M_{\odot}$. Comparing this figure to Fig. 3 of the 
Broadhurst et al. paper shows that unlike the best fit NFW profile this model 
exhibits the desired properties of more flattening towards the center $and$ 
more steepening towards the outside. The reduced $\chi^{2}$ statistic 
for this particular model fit is $\chi^{2}_{red}=23.8/dof=23.8/12=1.98$.

Figure 1 shows the run of gas entropy as it would be before any astrophysical 
processing. Note that both the general shape and the normalization are similar 
to the observations of Donahue et al. (2006).

\subsection{The Fornax Dwarf Spheroidal Galaxy}

Recently Goerdt et al. (2006) and S\'{a}nchez-Salcedo et al. (2006) 
have argued that the Fornax dwarf must contain a
large core in order that its globular clusters are not drawn into the 
center by dynamical friction. In order to test the sensitivity of this process 
to core size, two models were constructed. One has parameters ($1.36\times10^
{-5},0.385,0.1,1.92$) and $\beta_{max}=0.294$ while the other has ($3.46
\times10^{-6},0.920,1.31\times10^{-2},1.92$) and $\beta_{max}=0.237$. Both 
have the same virial mass of $1.99
\times10^{9} M_{\odot}$. (Note that the value of r$_{c}$ of 0.92 kpc provides 
a region within which the logarithmic density gradient is less than -0.1 of  
only $\sim0.2$ kpc consistent with Goerdt et al.)
The dynamical friction time scales can be compared 
using the following expression from Henon (1973) (i.e. $\tau_{df}\approx4
\times10^
{9}V^{3}/(ln(\Lambda)M_{GC}\rho)$ yrs). Here, V is the velocity of the cluster 
with assumed mass M$_{GC}=2\times10^{5} M_{\odot}$ and ln$\Lambda$ is the 
coulomb logarithm which we take here to be 5 for consistency with Goerdt et 
al. Given that within the radial distance $\sim0.2r_{c}$ the density and 
velocity dispersion are essentially constant and $\beta\sim0$, we replace 
V$^{3}/\rho$ with $3\sqrt{3}\sigma_{o}^{3}$/$\rho_{o}=3\sqrt{3}$/Q$_{o}$ and
obtain

\begin{equation}
\tau_{df}\approx\frac{5.2\times10^{9}}{M_{GC}Q_{o}}~~ yrs
\end{equation}

While the above estimate for $\tau_{df}$ is no substitute for a Goerdt et al. 
type of analysis, it does allow an intercomparison among our cored models. 

The above model with the largest core has a dynamical friction time scale of 
$7.5\times10^{9}$ years which is nearly four times that of the cluster with 
the smaller 
but higher density core. 

As a further check, a model with the same mass and r$_{c}$ but different 
$\alpha$ was made with parameters ($4.25\times10^{-6},0.920,1.49\times10^{-2},
1.0$). The
region within which the density gradient is less than -0.1 now has a radius of
only 0.05 
kpc. As expected, its dynamical friction time scale is shorter than the above 
model with the same r$_{c}$ but  only by $\sim18\%$. (Lowering $\alpha$ while 
keeping the other parameters the same lowers the central entropy slightly).
This simple analysis is in accord with the 
conclusions of Goerdt et al. and S\'{a}nchez-Salcedo et al. that increasing 
the core size leads to an increase in dynamical friction time scale. This 
increased time scale approaches a Hubble time and because our calculated 
value of $Q_{o}$ is $not$ necessarily assumed to be a result of phase packing 
remains 
within the constraints imposed by the sophisticated dynamical model of Fornax 
by Strigari et al. (2006). 

An additional check on the model comes from the 
recent 
work of Gilmore et al. (2007) whose analysis of the light distribution and 
velocity dispersion profile of several local dwarf spheroidal galaxies shows 
that shallow (cored) central density profiles with mean 
densities of $0.1~ M_{\odot}pc^{-3}$ (identical to that of our model above 
with r$_{c}=0.385~ kpc$) are most consistent with the observations.

\section{Discussion}

A new model for dark matter halos has been proposed and 
is successfully applied to observations of  
objects with masses ranging from $\sim10^{9}$ to $\sim10^{15}$
M$_{\odot}$.
Fig. 3 is a graphical summary of the results which shows 
the scaling relations discussed earlier. It is important to note key 
differences in 
structure occur with different scaling normalizations. The filled symbols are 
structures with relatively high values of $\rho_{o}$ 
($\sim 0.1 M_{\odot}~ pc^{-3}$) while the open 
symbols are structures with lower values of $\rho_{o}$ ($\sim 0.01 M_{\odot}
~ pc^{-3}$) and lower 
$\beta_{max}$'s. Close examination of Fig. 3 reveals a real 
systematic shift between these two groups of objects. Objects with identical 
central densities would lie essentially dispersionless along a line of the 
indicated slope. Further, for two models with 
the same mass, the one with the higher $Q_{o}$ has the higher $\rho_{o}$ and 
smaller $r_{c}$ (i.e. quantitatively $\partial$~log Q$_{o}$/$\partial$~log $
\rho_{o}=0.67$ and $\partial$~log Q$_{o}$/$\partial$~log r$_{c}=-1.57$ for a 
fixed halo mass). While the density is not expected to increase during merging,
Hernquist et al. (1993) propose a scheme whereby the density decreases while 
the dispersion remains constant. Dalcanton \& Hogan interpret this as a 
result of more violent merging so that different 
merging histories at earlier times could account for the variations in central 
density seen now. More observations are required to determine if the apparent 
dichotomy in central density is real and if so what its origin is. For example,
if we assume that the dichotomy extends to galaxy clusters, then a model with 
the same 
mass as A1689 but with one tenth the central density will have its gas entropy 
floor raised by $\sim2.8$. Such a change in central gas entropy floor is one 
characterization differentiating cooling flow clusters from non cooling flow 
clusters.

The error bars in each panel of Fig. 3 were calculated by assuming an 
empirical 
approximation to $\gamma$ (equation (4)) in terms of the central density (i.e.
$\gamma\sim1.80\rho^{-0.0753}_{o}$) and variables $\rho_{o}$ and r$_{c}$ were 
then treated as independent with estimated uncertainties of $\pm0.25$ and 
$\pm0.10$ respectively.

Rotation curves derived from the model and the NFW profile can be very similar
(e.g. the 
two A class galaxies in \S3.1) while according to the model the data appear
incompatible with standard CDM cosmological simulations which do not predict 
`observable-sized' 
cores. The similarity is most pronounced among objects in the high central 
density group. One property of our cored models which may be relevant to the 
`missing satellite' problem is that these structures are more vulnerable to 
tidal disruption than the NFW models, especially those 
halos with low central densities. 

From this work a relation has been determined between Q$_{o}$ and the mass 
of dark matter halos over a range from $\sim10^{9}$ to $\sim10^{15}$
M$_{\odot}$ . An observational challenge is to 
find the lower mass limit to objects with dark matter halos thus providing 
an estimate of the primordial value of Q (Q$_{p}$). Knowledge of 
Q$_{p}$ allows the determination of the mass of the dark matter particle 
(assuming that the particles are thermal) since then Q is proportional to the 
fourth power of the particle mass (e.g. equating the value of Q$_{o}$ found 
above for Fornax with Q$_{p}$ provides a lower limit on this mass of 431~ev). 
An additional constraint comes from an analysis of the power 
spectrum of the Ly$\alpha$ forest. From this one can 
determine the free streaming length ($\lambda_{fs}$) of the dark matter 
particle. This quantity in turn is simply related (in the case of thermal 
particles) to the particle mass. A recent determination of a limit on 
$\lambda_{fs}$ by
Seljak et al. (2006) implies a thermal dark matter particle mass limit of
$>10$ kev (i.e. Q$_{p}>1$ which according to our scaling relation above 
implies dark matter halos with masses as low as $\sim10^3$~M$_{\odot}$). 

An attractive alternative to the above `classical' WDM scenario has been 
proposed by Strigari et al. (2007). If the particles are non-thermally 
produced by the decay of a supersymmetric particle for example and if 
they are born sufficiently late then 
the initial velocities of the resulting daughter particles can be sufficiently 
high to yield a free streaming 
length comparable to that found from the Ly$\alpha$ forest analysis but with 
Q$_{p}$ orders of magnitude lower than above (i.e. Q$_{p}\sim10^{-5}-10^{-6}$)
and a correspondingly much higher dark halo mass limit. This picture has the 
additional feature that it is hierarchical in the conventional CDM sense since
the parent particles are born cold and being bosons they are not subject to 
the ultimate phase space density restriction.

Future results from experimental particle physics and even more sophisticated 
cosmological simulations should lead to a fuller understanding of the dark 
matter problem and the viability of the model.

\acknowledgements

The author wishes to thank Drs. Julio Navarro, Tony Burke and Andi Mahdavi 
for useful discussions, Dr. Greg Poole for introducing me to x-ray 
observations of galaxy clusters and the referee for a constructive report.

\clearpage

\appendix

\begin{center}
Appendix
\end{center}

As described above, a converged model is obtained by the somewhat 
cumbersome trial and error method. Below we present some analytical
approximations which will allow a more efficient exploration of parameter 
space and which illustrate more explicitly the two (scaling) parameter nature 
of the model (i.e. fix $\rho_{o}$ and vary Q$_{o}$). They were obtained by 
`fitting' to the converged 
solutions above. Letting $\alpha=1.92$ in equation (1) of the text,
the density profile is approximated by 

\begin{equation}
\rho(r)\sim\frac{\rho_{o}}{(1+(r/r_{c})^{1.92})(1+(r/(\delta(\rho_{o})r_{c}))
^{2.08})}
\end{equation}

where

\begin{equation}
r_{c}\sim5.773\times 10^{-3}~\rho_{o}^{-0.2043}Q_{o}^{-1/3}
\end{equation}

and

\begin{equation}
\delta(\rho_{o})\sim6.93~log~\rho_{o}+24.97
\end{equation}

The velocity dispersion becomes 
\begin{equation}
\sigma^{2}(r)\sim\frac{(\rho_{o}/Q_{o})^{2/3}}{(1+(r/(\delta(\rho_{o})r_{c}))^
{2.08})^{2/3}}
\end{equation}

Finally the value of $\beta_{max}$ is obtained by determining the maximum 
associated with the smallest r value (i.e. the first maximum) in the following
expression

\begin{equation}
\beta(r)\sim-0.5GM_{r}/(\sigma^{2}r)-0.5~d~log~Q/d~log~r-(5/6)d~log~\sigma^{3}/
d~log~r
\end{equation}

where M$_{r}$ comes from the integration of equation (3) in the text.

With these approximations, values of r$_{c}$, M$_{100}$, R$_{100}$, are 
determined to $<1\%$, $<4\%$, and $<2\%$.  Deviations in $\beta_{max}$ are 
between $1.4\%$ 
and $14.1\%$ with the largest deviations occurring at the lowest values of 
$\rho_{o}$. Beyond the maxima the run of $\beta$ is not reliable. Circular 
velocity maxima derived from the above expressions are 
within 5$\%$ of the model values. It should be emphasized that the above 
expressions and bounds were determined 
from models with central densities $5.7\times 10^{-3}\leq\rho_{o}\leq2.54
\times 10^{-1}$.

\clearpage

\begin{figure}
\epsscale{.80}
\plotone{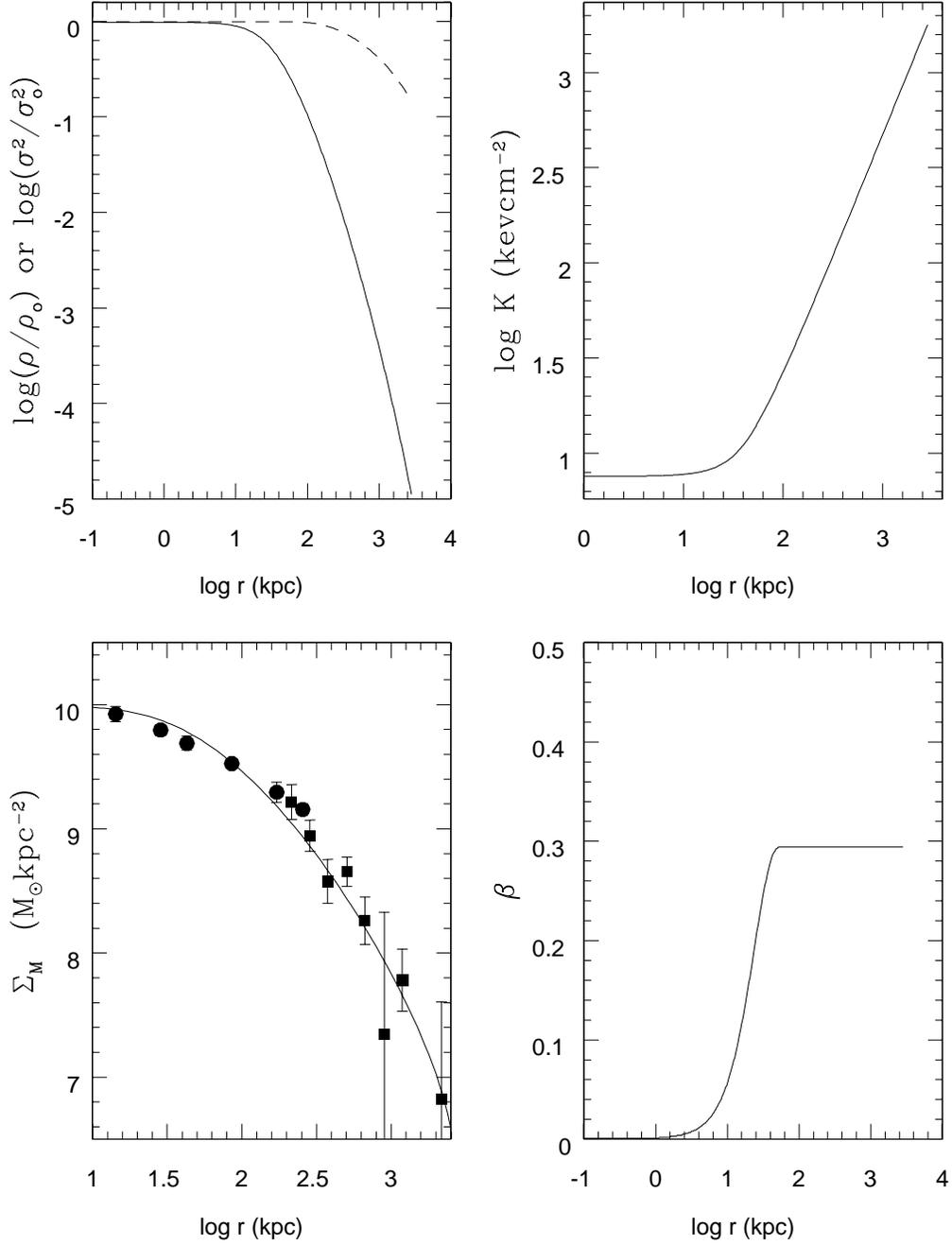}
\caption{Attributes of a solution to equations (1),(2),\& (3) for the galaxy 
cluster A1689. Upper left panel: The run of density (solid) and velocity 
dispersion (dashed) versus radial distance. Upper right panel: The gas 
entropy profile before astrophysical processes change it. Lower left panel: 
Observational data from Fig. 3 of Broadhurst et al. 2005 with the model 
projected mass density superposed. Lower right panel: The run of $\beta$ 
with radial distance derived as described in \S2. Log(r$_{c}$) for this model 
is 1.52.} 
\end{figure}

\clearpage

\begin{figure}
\plotone{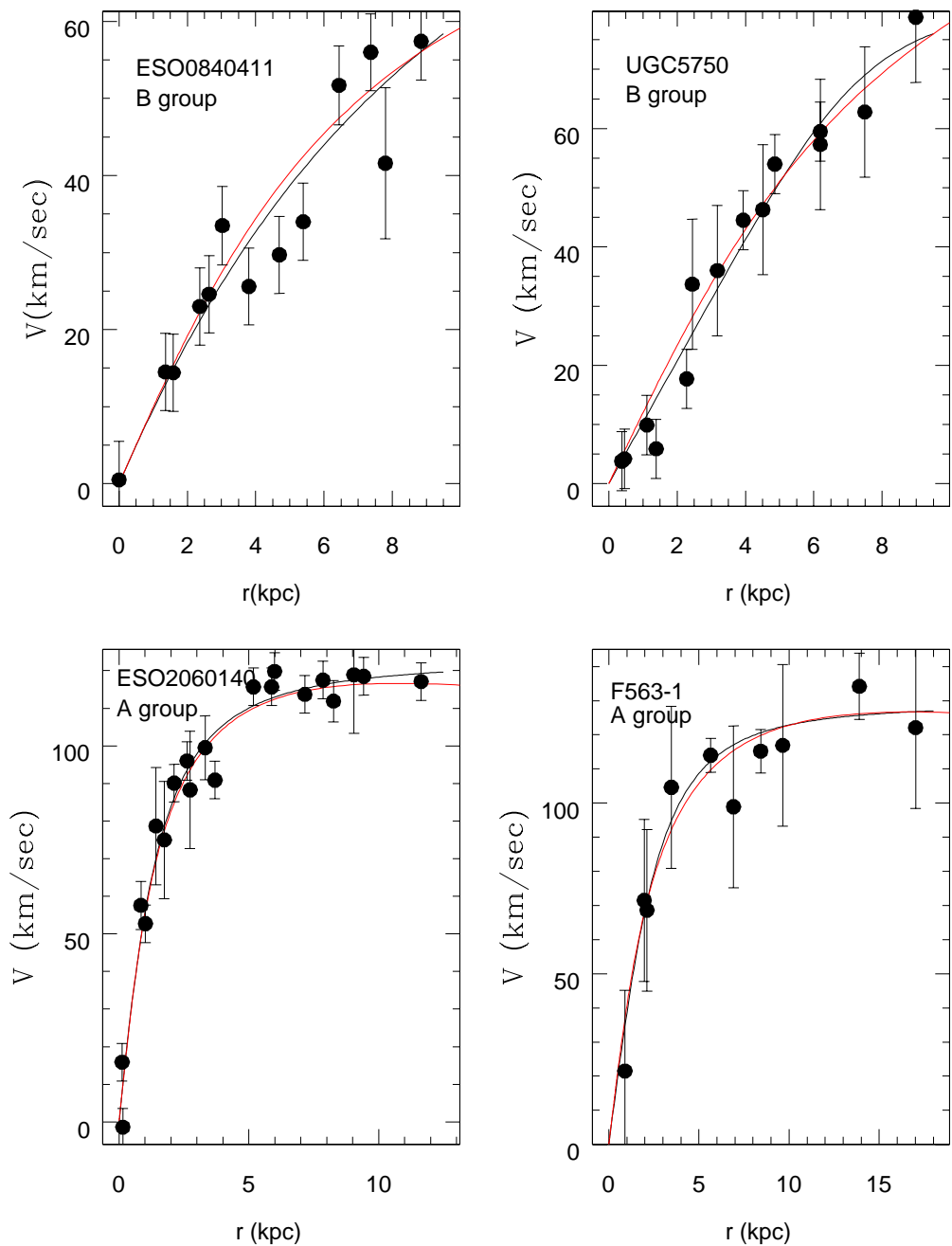}
\caption{Model fits to LSB galaxy rotation curves. The black line is the fit 
of an empirical fitting function with the same parameters given in Figs 7 \& 
8 of Hayashi et al. 2004. The red curve is the model fit. Recall that NFW 
profiles could not be well fit to the two group B galaxies. The reduced 
$\chi^{2}$ for all model fits is $<1$.}
\end{figure}

\clearpage

\begin{figure}
\plotone{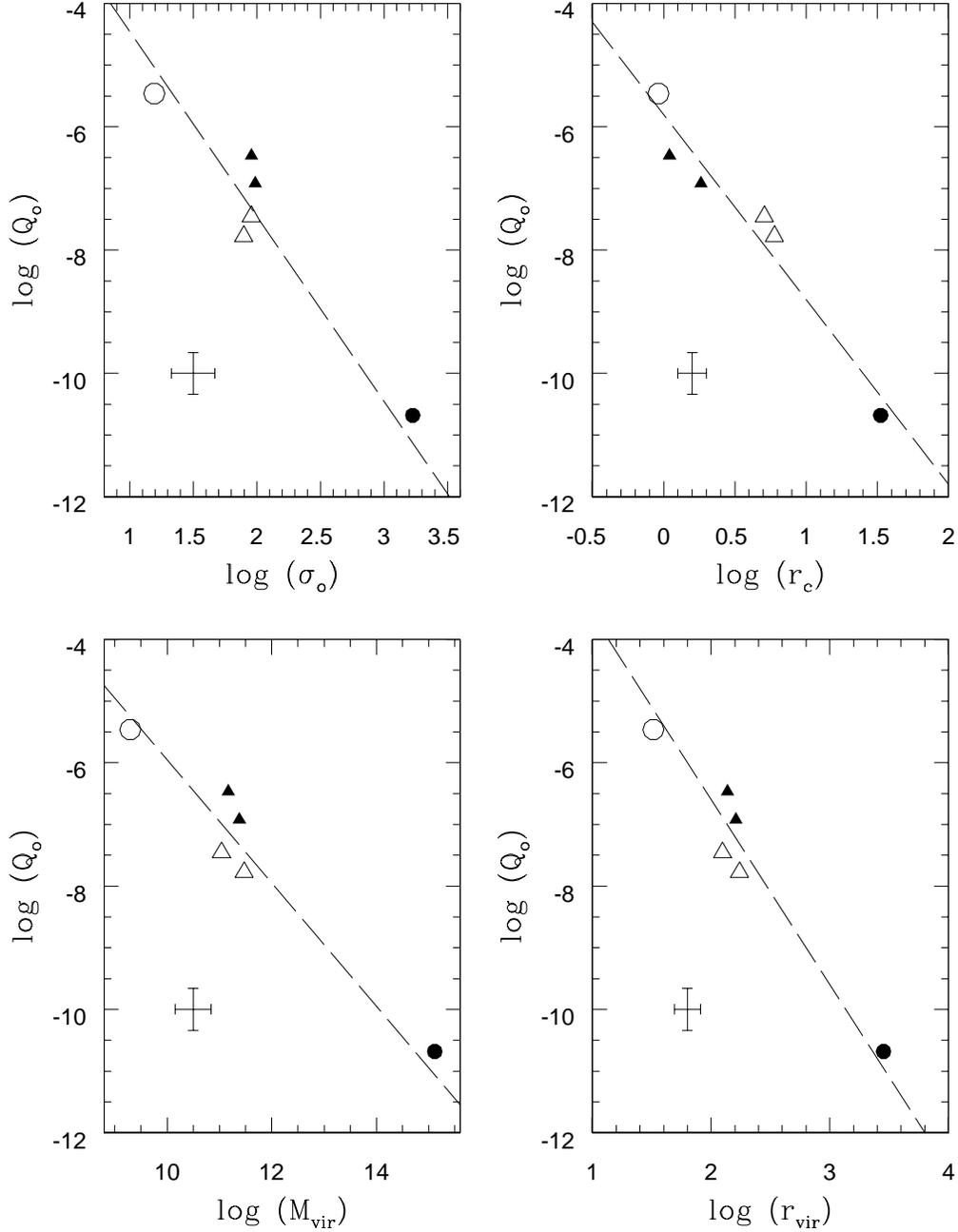}
\caption{The model parameters for the six systems discussed here. Open circle-
Fornax dwarf, open triangles-group B LSB galaxies, solid triangles-group A 
LSB galaxies and solid circle-galaxy cluster A1689. Clockwise from the upper 
left shows log($Q_{o}$) versus central velocity dispersion, core radius, 
virial radius and virial mass. The dashed lines are not fits but illustrate 
the scaling relations described in the text (i.e. $Q_{o}\propto \sigma_{o}^{-3}
$, $\propto r_{c}^{-3}$, $\propto r_{vir}^{-3}$ and $\propto M_{vir}^{-1}$).}
\end{figure}

\end{document}